\documentclass{article}
\usepackage{arxiv}
\usepackage[utf8]{inputenc}
\usepackage[T1]{fontenc}
\usepackage[inline]{enumitem}
\usepackage{booktabs}
\usepackage{amsmath}
\usepackage{amsfonts}
\usepackage{amssymb}
\usepackage{enumitem}
\usepackage{mathbbol}
\usepackage{mathtools}
\usepackage{hyperref}
\usepackage{url}
\usepackage{nicefrac}
\usepackage{microtype}
\usepackage{natbib}
\usepackage{multirow}

\newcommand{\eg}{\emph{e.g.}}
\newcommand{\wrt}{\emph{w.r.t. }}

\title{Duet at TREC 2019 Deep Learning Track}

\author{
  Bhaskar Mitra \\
  Microsoft, University College London \\
  Montreal, Canada \\
  \texttt{bmitra@microsoft.com} \\
   \And
 Nick Craswell \\
  Microsoft \\
  Redmond, USA \\
  \texttt{nickcr@microsoft.com} \\
}

\begin{document}
\maketitle

\begin{abstract}
This report discusses three submissions based on the Duet architecture to the Deep Learning track at TREC 2019.
For the document retrieval task, we adapt the Duet model to ingest a ``multiple field'' view of documents---we refer to the new architecture as Duet with Multiple Fields (DuetMF).
A second submission combines the DuetMF model with other neural and traditional relevance estimators in a learning-to-rank framework and achieves improved performance over the DuetMF baseline.
For the passage retrieval task, we submit a single run based on an ensemble of eight Duet models.
\end{abstract}


\section{Introduction}
\label{sec:intro}

The Duet architecture was proposed by \citet{mitra2017learning} for document ranking.
Fig. 7 from The original paper show that the retrieval effectiveness of the model is still improving as the size of the training data approaches $2^{17}$ samples.
The training data employed in that paper is a proprietary dataset from Bing.
A similar plot was later reproduced on a public benchmark by \citet{nanni2017benchmark}, but in the context of a passage ranking dataset with synthetic queries.
Variations of the Duet model \citep{mitra2019updated,  mitra2019incorporating, cohen2018cross} have since then been evaluated on other public passage ranking datasets.
However, the lack of large scale training data prevented the public evaluation of Duet for document ranking.

The deep learning track at TREC 2019 makes large training datasets---suitable for traininig deep models with large number of learnable parameters---publicly available in the context of a document ranking and a passage ranking tasks.
We benchmark the Duet model on both tasks.

In the context of the document ranking task, we adapt the Duet model to ingest a ``multiple field'' view of the documents, based on findings from \citet{zamani2018neural2}.
We refer to this new architecture as Duet with Multiple Fields (DuetMF) in the paper.
Furthermore, we combine the relevance estimates from DuetMF with several other traditional and neural retrieval methods in a learning-to-rank (LTR) \citep{Liu:2009} framework.

For the passage ranking task, we submit a single run based on an ensemble of eight Duet models.
The architecture and the training scheme resembles that of the ``Duet V2 (Ensembled)'' baseline listed on the MS MARCO leaderboard\footnote{\url{http://www.msmarco.org/leaders.aspx}}.

\section{TREC 2019 deep learning track}
\label{sec:task}
The TREC 2019 deep learning track introduces:
\begin{enumerate*}[label=(\roman*)]
    \item a document retrieval task and
    \item a passage retrieval task.
\end{enumerate*}
For both tasks, participants are provided a set of candidates---$100$ documents and $1000$ passages, respectively---per query that should be ranked.
Participants can choose to either rerank provided candidates or retrieve from the full collection.

For the passage retrieval task, the track reuses the set of $500$K+ manually-assessed binary training labels released as part of the Microsoft Machine Reading COmprehension (MS MARCO) challenge \citep{bajaj2016ms}.
For the document retrieval task, the passage-level labels are transferred to their corresponding source documents---producing a training dataset of size close to 400K labels.

For evaluation, a shared test set of $200$ queries is provided for both tasks, of which two different overlapping set of $43$ queries were later selected for manual NIST assessments corresponding to the two tasks.

Full details of all datasets is available on the track website\footnote{\url{https://microsoft.github.io/TREC-2019-Deep-Learning/}} and in the track overview paper \citep{craswell2019overview}.

\section{Methods and results}
\label{sec:runs}

The Duet model proposed by \citet{mitra2017learning} employs two deep neural networks trained jointly towards a retrieval task:
\begin{enumerate*}[label=(\roman*)]
    \item The ``distributed'' sub-model learns useful representations of text for matching and
    \item the ``local'' sub-model estimates relevance based on patterns of exact term matches between query and document.
\end{enumerate*}
\citet{mitra2019updated} propose several modifications to the original Duet model that show improved performance on the MS MARCO passage ranking challenge.
We adopt the updated Duet model from \citet{mitra2019updated} and incorporate additional modifications, in particular to consider multiple fields for the document retrieval task.
Table \ref{tbl:results} summarizes the official evaluation results for all three runs.

\begin{table}
    \small
    \centering
    \caption{Official TREC results.
    The recall metric is computed at position 100 for the document retrieval task and at position 1000 for the passage retrieval task.}
    \begin{tabular}{lllllccccc}
    \hline
    \hline
        \textbf{Run description} & \textbf{Run ID} & \textbf{Subtask} & \textbf{MRR} & \textbf{NDCG@10} & \textbf{MAP} & \textbf{Recall} \\
        \hline
        \textbf{Document retrieval task} \\
        LTR w/ DuetMF as feature & ms\_ensemble & fullrank & 0.876 & 0.578 & 0.237 & 0.368 \\
        DuetMF model & ms\_duet & rerank & 0.810 & 0.533 & 0.229 & 0.387 \\
        \hline
        \textbf{Passage retrieval task} \\
        Ensemble of 8 Duet models & ms\_duet\_passage & rerank & 0.806 & 0.614 & 0.348 & 0.694 \\
        \hline
        \hline
    \end{tabular}
    \label{tbl:results}
\end{table}

\paragraph{Duet model with Multiple Fields (DuetMF) for document ranking.}

\begin{figure}
    \centering
    \includegraphics[width=0.95\textwidth]{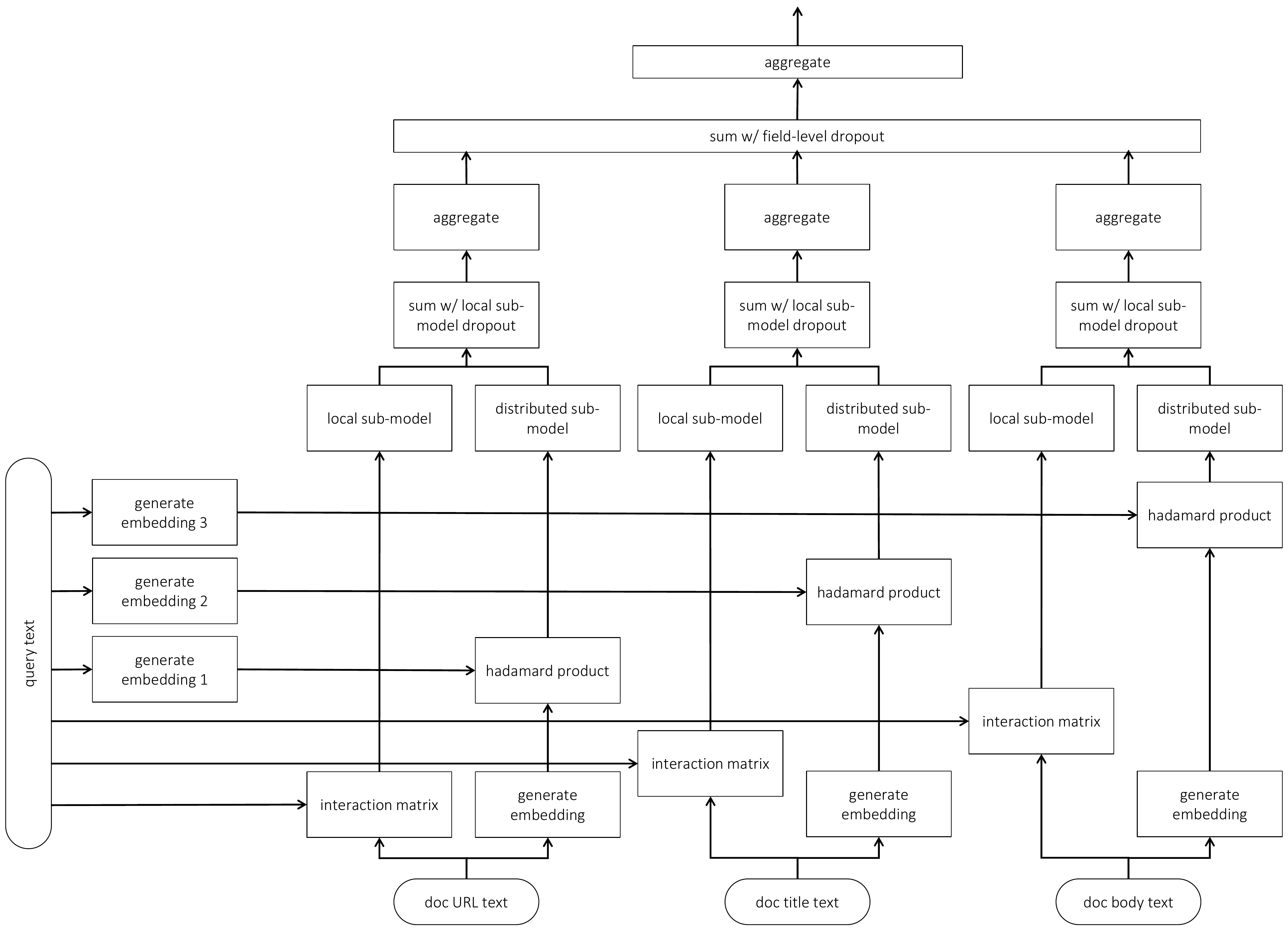}
    \caption{The modified Duet model (DuetMF) that considers multiple document fields.}
    \label{fig:model}
\end{figure}

\citet{zamani2018neural2} study neural ranking models in the context of documents with multiple fields.
In particular, they make the following observations:
\begin{enumerate}[label=Obs. \arabic*:]
    \item It is more effective to summarize the match between query and individual document fields by a vector---as opposed to a single score---before aggregating to estimate full document relevance to the query.
    \item It is better to learn different query representations corresponding to each document field under consideration.
    \item Structured dropout (\eg, field-level dropout) is effective for regularization during training.
\end{enumerate}
We incorporate all of these ideas to modify the Duet model from \citet{mitra2019updated}.
The updated model is shown in Fig. \ref{fig:model}.

Documents in the deep learning track dataset contains three text fields:
\begin{enumerate*}[label=(\roman*)]
    \item URL,
    \item title, and
    \item body.
\end{enumerate*}
We employ the Duet architecture to match the query against each individual document fields.
In line with Obs. 1 from \citep{zamani2018neural2}, the field-specific Duet architecture outputs a vector instead of a single score.
We do not share the parameters of the Duet architectures between the field-specific instances based on Obs. 2.
Following Obs. 3, we introduce structured dropouts at different stages of the model.
We randomly dropout each of the local sub-models for $50\%$ of the training samples.
Similarly, we also dropout different combinations of field-level models uniformly at random---taking care that at least one field-level model is always retained.

We consider the first $20$ terms for queries and for document URLs and titles.
For document body text, we consider the first $2000$ terms.
Similar to \citet{mitra2019updated}, we employ pretrained word embeddings as the input text representation for the distributed sub-models.
We train the word embeddings using a standard word2vec \citep{mikolov2013distributed} implementation in FastText \citep{joulin2016bag} on a combination of the MS MARCO document corpus and training queries.

Similar to previous work \citep{mitra2017learning, mitra2019updated}, the query and document field embeddings are learned by deep convolutional-pooling layers.
We set the hidden layer size at all stages of the model to $300$ and dropout rate for different layers to $0.5$.
For training, we employ the RankNet loss \citep{burges2005learning} over $<q, d_\text{pos}, d_\text{neg}>$ triples and the Adam optimizer \citep{kingma2014adam}---with a minibatch size of $128$ and a learning rate of $0.0001$ for training.
We sample $d_\text{neg}$ uniformly at random from the top $100$ candidates provided that are not positively labeled.
When employing structured dropout, the same sub-models are masked for both $d_\text{pos}$ and $d_\text{neg}$.

In light of the recent success of large pretrained language models---\eg, \citep{nogueira2019passage}---we also experiment with an unsupervised pretraining scheme using the MS MARCO document collection.
The pretraining is performed over $<q_\text{pseudo}, d_\text{pos}, d_\text{neg}>$---where $d_\text{pos}$ and $d_\text{neg}$ are randomly sampled from the collection and a pseudo-query $q_\text{pseudo}$ is generated by picking the URL or the title of $d_\text{pos}$ randomly (with equal probability) and masking the corresponding field on the document side for both $d_\text{pos}$ and $d_\text{neg}$.
We see faster convergence during supervised training when the DuetMF model is pretrained in this fashion on the MS MARCO document collection.
We posit that a more formal study should be performed in the future on pretraining Duet models on large collections, such as Wikipedia and the BookCorpus \citep{zhu2015aligning}.

\paragraph{Learning-to-rank model for document ranking.}
We train a neural LTR model with two hidden layers---each with $1024$ hidden nodes.
The LTR run reranks a set of $100$ document candidates retrieved by query likelihood (QL) \citep{ponte1998language} with Dirichlet smoothing ($\mu=1250$) \citep{mackay1995hierarchical}.
Several ranking algorithms based on neural and inference networks act as features:
\begin{enumerate*}[label=(\roman*)]
    \item DuetMF,
    \item Sequential Dependence Model (SDM) \citep{metzler2005markov}, and
    \item Pseudo-Relevance Feedback (PRF) \citep{lavrenko2001relevance, lavrenko2008generative},
    \item BM25, \citep{robertson2009probabilistic}, and
    \item Dual Embedding Space Model (DESM) \citep{nalisnick2016improving, mitra2016desm}.
\end{enumerate*}

We employ SDM with an order of $3$, combine weight  of $0.90$, ordered window weight of $0.034$, and an unordered window weight of $0.066$ as our base candidate scoring function.
We use these parameters to retrieve from the target corpus as well as auxiliary corpora of English language Wikipedia ({\tt enwiki-20180901-pages-articles-multistream.xml.bz2}), LDC Gigaword ({\tt LDC2011T07}).
For PRF, initial retrievals---from either of the target, wikipedia, or gigaword corpora---adopted the SDM parameters above, however are used to rank 75-word passages with a 25-word overlap.
These passages are then interpolated using the top $m$ passages and standard relevance modeling techniques, from which we select the top $50$ words to use as an expanded query for the final ranking of the target candidates.
We do not explicitly adopt RM3 \citep{abdul2004umass} because our LTR model implicitly combines our initial retrieval score and score from the expanded query.
All code for the SDM and PRF feature computation is available at \url{https://github.com/diazf/indri}.

We evaluate two different BM25 models with hyperparameters $<k_1=0.9, b=0.4>$ and $<k_1=3.44, b=0.87>$.

Corresponding to each of the DuetMF, SDM, PRF, and BM25 runs we generate two features based on the score and the rank that the model predicts for a document \wrt the target query.

We generate eight features by comparing the query against two different document fields (title and body) and using different DESM similarity estimates (INxIN, INxOUT, OUTxIN, OUTxOUT).

Lastly, we add couple of features based on query length and domain quality---where the latter is defined simply as a ratio between how often documents from a given domain appear in the positively labeled training data and in the overall document collection.

\paragraph{Ensemble of Duet models for passage ranking.}
For the passage ranking task, we adopt the exact same model and training procedure from \citep{mitra2019updated}.
Our final submission is an ensemble of eight Duet models.

\section{Discussion and conclusion}
\label{sec:conclusion}

One of the main goals of the deep learning track is to create a public reusable dataset for benchmarking the growing body of neural information retrieval literature \citep{mitra2018introduction}.
We submit three runs based on the Duet architecture for the two---document and passage---retrieval tasks.
Our main goal is to enrich the set of pooled documents for NIST assessments with documents that a Duet based architecture is likely to rank highly.
As a secondary goal, we are also interested in benchmarking Duet against other state-of-the-art neural and traditional methods.
A more detailed comparison of the performance of these Duet runs with other TREC submissions is provided in the track overview paper \citep{craswell2019overview}.

\bibliographystyle{plainnat}
\bibliography{bibtex}

\end{document}